\documentclass[aps,prd,showpacs,nofootinbib]{revtex4-1}
\usepackage{graphicx,amsfonts,amssymb,amsbsy}
\usepackage{natbib}
\usepackage{pstricks}
\usepackage{amsmath}
\usepackage{xcolor}
\usepackage{color}

\newcommand{\beq}{\begin{equation}}   
\newcommand{\eeq}{\end{equation}}

\newcommand{\Eq}[1]{{Eq.~({\ref{#1}})}}

\begin{document}

\title{Crystalline chiral condensates as a component of compact stars }
\author{S. Carignano}
\affiliation{Department of Physics, University of Texas at El Paso,
El Paso, TX 79968, USA}
\author{E. J. Ferrer}
\affiliation{Department of Physics, University of Texas at El Paso,
El Paso, TX 79968, USA}
\author{V. de la Incera}
\affiliation{Department of Physics, University of Texas at El Paso,
El Paso, TX 79968, USA}
\author{L. Paulucci}
\affiliation{Universidade Federal do ABC, Rua Santa Ad\'elia, 166, 09210-170 Santo Andr\'e, SP, Brazil}

\begin{abstract}

\noindent
We investigate the influence of spatially
inhomogeneous chiral symmetry-breaking condensates
 in a magnetic field background on the equation of state for compact stellar objects. After building a hybrid star composed of nuclear and quark matter using the Maxwell construction, we find, by solving the Tolman-Oppenheimer-Volkoff equations for stellar equilibrium, that our equation of state supports stars with masses around 2 $M_\odot$ for values of the magnetic field that are in accordance with those inferred from magnetar data. The inclusion of a weak vector interaction term in the quark part allows one to reach 2 solar masses for relatively small central magnetic fields, making this composition a viable possibility for describing the internal degrees of freedom of this class of astrophysical objects. 
\end{abstract}


\maketitle

\section{Introduction}

\noindent
The study of the properties of hadronic matter under extreme conditions 
is perhaps the most challenging task in contemporary nuclear physics. 
Due to its non-perturbative behavior, quantum chromodynamics (QCD), the fundamental theory of strong interactions, cannot be solved using conventional field-theoretical methods, making any prediction of its properties  an extremely challenging task.
While \textit{ab-initio} lattice calculations and current heavy-ion experiments at the RHIC and LHC  shed some light on the properties of strongly interacting matter at high temperatures, the opposite region of the QCD phase diagram, associated to high density and low temperature conditions, 
is still largely unknown. On the theoretical side, lattice calculations at nonzero chemical potentials are hindered by the sign problem and most of the current predictions rely on phenomenological models, while experimentally none of the current heavy-ion colliders can reach sufficient densities to probe this region.

In spite of the current uncertainties, the phase structure of QCD at finite densities is nevertheless expected to be very rich (see e.g. \cite{Fukushima:2013rx}). In particular, a growing consensus has been recently building around the idea that crystalline phases might appear in the intermediate density region (up to a few times nuclear matter density), before the onset of color superconductivity. The formation of such phases could in principle delay the restoration of chiral symmetry, dramatically altering the properties of cold quark matter (for a recent review, see \cite{Buballa:2014tba}). In particular, it has been suggested that the presence of strong background magnetic fields, a natural element in astrophysical scenarios, might significantly enhance the window for inhomogeneous phases \cite{Klimenko2010,Tatsumi:2014}, possibly leading to significant effects in the equation of state (EoS) of dense quark matter, as we discuss below.

 While waiting for the next generation of heavy-ion colliders such as the FAIR experiment in Darmstadt and NICA in Dubna, which promise to access experimentally this window of the QCD phase diagram, the best laboratory for investigating properties of dense matter is given by compact stellar objects, which provide the only known realization of ultradense systems in nature. 
Of particular interest are measurements of masses and radii, which indirectly provide numerous hints on the possible EoS for QCD at high densities.
In particular, the recent discoveries of stars with masses close to 
$2M_{\odot}$ ($M_{\odot}$ being the solar mass) \cite{Demorest,Antoniadis}
impose rather strong limitations on the possible EoS.
It is worth recalling that a lot of modeling of the microscopic physics inside these compact stellar objects is involved when making this kind of prediction, introducing a large number of uncertainties. Nevertheless, while the role of hyperons as a softening ingredient of the nuclear EoS is still under debate \cite{hyper1, *hyper2, *hyper3, *hyper4, *hyper5, *Bednarek, BlaschkeEVA, Baldo}, as well as the presence of quark matter \cite{QM1, *QM2, QM3}, numerous studies seem to indicate that most of the ordinary phases of (confined or deconfined) matter might not be able to support the large stellar masses observed. All these considerations suggest that some fundamental aspect in the physics of dense matter might still be missing from these calculations. Due to the high densities reached in the core of compact stellar objects, it might be reasonable to expect a transition to more exotic phases, whose EoS could be stiff enough to sustain these massive stars.
Of course, the issue of the maximum mass is  subject to other effects as well, such as high rotation rates (see e.g. \cite{Weber} and references therein) or the existence of strong magnetic fields \cite{mag1, *mag2, H_profile, mag3, Sotani} that affect the EoS and may allow those objects to support higher masses than a static, nonmagnetized star would. 

The main purpose of this work is to investigate the effects of the formation of inhomogeneous chiral-symmetry-breaking condensates in a magnetic field background on the EoS of cold and dense matter, and whether they can lead to predictions for compact stellar objects which are compatible with current experimental observations. 
In particular, we aim at building hybrid stars with a crystalline quark matter core, using the resulting EoS as input for the Tolman-Oppenheimer-Volkoff (TOV) equations. In order to build a realistic description of matter for astrophysical scenarios, the models under consideration will include the effects of strong magnetic fields, which are naturally expected to be present in the compact stellar medium.

This work is structured as follows: in Secs. II-III, we introduce the phenomenological models employed to describe quark and nuclear matter. In Sec. IV, we consider a medium-dependent magnetic field and give its density profile for various parametrizations. In Sec. V, we build the EoS for a hybrid star with a crystalline quark matter core in the presence of such magnetic field, and obtain the corresponding mass-radius (M-R) plots. Finally in Sec. VI we summarize the results and give our concluding remarks.

\section{Models of Neutral Magnetized Quark Matter with Inhomogeneous Condensates}
\noindent
In this section we introduce the models we are going to use to investigate quark matter with spatially inhomogeneous chiral condensates in a magnetic field background. Since our ultimate goal is to investigate the structure of compact stars,
 we shall impose the physical conditions of electrical neutrality and $\beta-$equilibrium in all our calculations.  Vector interactions will also be included.
For our calculations, we will consider both two- and three-flavor models, which will be described in the following.

\subsection{Two-flavor model}
\noindent
To study two-flavor quark matter in a magnetic field background we consider the following Nambu--Jona-Lasinio (NJL)-type Lagrangian density
\begin{equation} \label{L-2fl}
\mathcal{L}^{(2f)} =\bar\psi \left(i \gamma^\mu D_\mu + \mu \gamma^{0} - m_q\right) \psi +\bar{\psi}_e \left(i \gamma^\mu D^{(e)}_\mu - m_e \right)\psi_e+\mathcal{L}_{int} \,,
\end{equation}
containing a doublet of quark fields $\psi^T= (\psi_u,\psi_d)$ in flavor space, with current mass matrix $m_q=\textrm{diag} (m_u,m_d)$ and an electron field $\psi_e$ of mass $m_e$. A nonzero baryon density has been introduced via the quark baryon chemical potential $\mu$. 
The covariant derivative describing the coupling of matter with a static external magnetic field along the $z$-direction is  $D_\mu=\partial_\mu+iQA^{ext}_\mu$, with electric charge matrix in flavor space $Q=\textrm{diag}(e_u,e_d) = \textrm{diag}(\frac{2}{3}e,-\frac{1}{3}e)$, $e$ being the unit electric charge and $A^{ext}_\mu$ being the external electromagnetic four-potential taken in the Landau gauge $A^{ext}_\mu= (0,0,Hx,0)$. 

The quark interaction Lagrangian  $\mathcal{L}_{int}$ is given by
\begin{equation} \label{Lint}
\mathcal{L}_{int} =\mathcal{L}_{1}+\mathcal{L}_{2}+\mathcal{L}_{V} \,,
\end{equation}
with
\begin{equation}\label{l-1}
\mathcal{L}_1 =G_1 \left[ (\bar\psi\psi)^2 + (\bar\psi i\gamma^5\psi)^2 + (\bar\psi\tau^a\psi)^2 + (\bar\psi i\gamma^5 \tau^a \psi)^2 \right],
\end{equation}
\begin{equation}\label{L2}
\mathcal{L}_2 =G_2 \left[ (\bar\psi\psi)^2 - (\bar\psi i\gamma^5\psi)^2 - (\bar\psi\tau^a\psi)^2 + (\bar\psi i\gamma^5 \tau^a \psi)^2 \right], 
\end{equation}
and vector channel
\begin{equation}\label{l-V}
\mathcal{L}_V =-G_V \left[ (\bar\psi \gamma_\mu \psi)^2 + (\bar\psi \gamma^5\gamma_\mu\psi)^2 + (\bar\psi\gamma_\mu\tau^a\psi)^2 + (\bar\psi \gamma^5\gamma_\mu \tau^a \psi)^2 \right] \,.
\end{equation}
Here, the matrices $\tau_a$ are the generators of the SU(2) flavor group and a sum in the color index is assumed in all the quark terms.

For applications to compact stellar objects, one needs to consider electrically neutral and $\beta$-equilibrated matter. To incorporate neutrality, we insert an electric charge term $-\mu_e \mathbb{Q}_e$ in (\ref{L-2fl}), with charge operator 
\begin{equation}\label{elctricQ2f}
\mathbb{Q}_e =\frac{2}{3}\bar\psi_u\gamma_0 \psi_u - \frac{1}{3}\bar\psi_d\gamma_0 \psi_d-\bar\psi_e\gamma_0 \psi_e.
\end{equation}
The electric chemical potential $\mu_e$ is not an independent parameter; it has to be determined self-consistently from the condition of electrical neutrality. A nonzero $\mu_e$ gives rise to an isospin asymmetry between the quarks, which now have chemical potentials
\beq
\label{eq:splitmu}
\mu_u = \mu - \frac{2}{3} \mu_e \,, \qquad \mu_d = \mu + \frac{1}{3} \mu_e \,.
\eeq

At this point we perform the standard mean-field approximation and introduce 
an inhomogeneous ansatz for the expectation values $\langle\bar\psi_f\psi_f\rangle$ and $\langle\bar\psi_f i\gamma_5\psi_f\rangle$, $f=u,d$ 
(in the following we neglect flavor off-diagonal mean fields corresponding to charged pion condensation).  
We note that the interaction term  (\ref{L2}), which corresponds to the instanton contribution $G_2[\det\bar\psi (1+\gamma_5)\psi+\det\bar\psi (1 -\gamma_5)\psi]$ \cite{Asakawa,Klevansky} and  is added to the Lagrangian to include the $U(1)_A$ anomaly of QCD, contains flavor mixing terms like $\langle\bar\psi_u\psi_u\bar\psi_d\psi_d\rangle$ that would significantly complicate our calculation when dealing with asymmetric inhomogeneous matter.
 As a first step, we then choose to avoid dealing with mixing terms by  considering a simpler version of our model in which different quark flavors can be completely decoupled by neglecting the instanton term in the interaction Lagrangian (i.e. taking $G_2=0$). We expect the main features of the quark EoS to be qualitatively unaffected by this simplification, and note that at any rate the quark condensates will still influence each other through the neutrality condition. From now on, we will consider $G_1=G_S$ and work in the chiral limit $m_u = m_d = 0$.

Let us consider the following plane-wave ansatz
\begin{align}
 \label{condensates2f}
-4G_S\langle\bar\psi_u\psi_u\rangle & =\Delta_u \cos (q_u z) \,,  \quad -4G_S\langle\bar\psi_u i\gamma_5\psi_u\rangle = \Delta_u \sin (q_u z) \,, \nonumber
\\
-4G_S\langle\bar\psi_d\psi_d\rangle & = \Delta_d \cos (q_d z) \,, \quad -4G_S\langle\bar\psi_d i\gamma_5\psi_d\rangle =\Delta_d \sin (q_d z) \,.
 \end{align}
In the isospin-symmetric case, this ansatz reduces to the so-called ``chiral density wave'' (CDW) \cite{Nakano:2005}, characterized by plane-wave condensates with magnitude $\Delta=\Delta_u=\Delta_d$ for each flavor and equal and opposite wave vectors, $q=q_u=-q_d$. However, in isospin-asymmetric matter, there is no reason why different flavors should have the same condensate, and one should allow for two separate amplitudes $\Delta_f$ and modulations $q_f, f=u,d$.

\noindent In an external magnetic field, the only rotational symmetries that survive are the SO(2) group of rotations about the field direction, making that direction special. That is why we have chosen the modulation of the condensates along the magnetic field direction.  We note that a modulation along the field direction is known to be energetically favored in the isospin-symmetric case \cite{Nakano:2005}.  

Since we are working at nonzero baryon density in a theory with vector interactions, we also introduce expectation values of the individual quark number densities for each flavor,
\beq
\label{quarkdensities}
\langle\bar\psi_u\gamma_0\psi_u\rangle= \rho_u, \quad \langle\bar\psi_d\gamma_0\psi_d\rangle= \rho_d, 
\eeq
connected to the baryon density through $3 \rho_B= (\rho_u+\rho_d)$. While for an arbitrary spatial dependence of the chiral condensate a proper self-consistent inclusion of the expectation values of the quark number densities could be challenging, as they could be themselves inhomogeneous, in the case of a CDW modulation the procedure is actually straightforward, thanks to the fact that, for this particular ansatz, the quark number density is spatially constant \cite{CNB:2010}. The net effect of including vector interactions then amounts, just like for homogeneous matter, to the introduction of a shifted chemical potential for each flavor, given by \cite{Zhang:2009,Abuki:2009}
\beq
\label{eq:splitmu+v}
\widetilde{\mu}_u = \mu_u -4G_V \rho_u   \,, \qquad \widetilde{\mu}_d = \mu_d-4G_V \rho_d \,.
\eeq

Expanding around the expectation values introduced in Eqs. (\ref{condensates2f}) and (\ref{eq:splitmu+v}), we obtain the mean-field Lagrangian 
\begin{eqnarray} \label{L-2flMF}
\mathcal{L}^{(2f)}_{MF} &=&\bar\psi_u \left(i \gamma^\mu D^{(u)}_\mu + \widetilde{\mu}_u \gamma^{0} -\Delta_{u}e^{i\gamma_5 q_uz}\right) \psi_u +
\bar\psi_d \left(i \gamma^\mu D^{(d)}_\mu + \widetilde{\mu}_d \gamma^{0} -\Delta_{d}e^{i\gamma_5 q_dz}\right) \psi_d \nonumber
\\
&+&\bar{\psi}_e \left(i \gamma^\mu D^{(e)}_\mu+\mu_e\gamma^{0} - m_e \right)\psi_e -\frac{\Delta^2_u}{8G_S}-\frac{\Delta^2_d}{8G_S}+\frac{(\widetilde{\mu}_u-\mu_u)^2}{8G_V} +\frac{(\widetilde{\mu}_d-\mu_d)^2}{8G_V}  \,.
\end{eqnarray}

\noindent
Since this Lagrangian is now bilinear in the matter fields, the corresponding thermodynamic potential can be readily obtained. In the following, we neglect thermal effects and work at zero temperature, a reasonable approximation when describing cold and dense stellar matter. The zero-temperature, mean-field thermodynamic potential of the two-flavor theory (\ref{L-2flMF}) is thus given by 
\beq
\label{eq:omega2f}
 \Omega^{(2f)}=\Omega_e + N_c \sum_{f=u,d} \Omega_f + \sum_{f=u,d}\left[\frac{\Delta^2_f}{8G_S}-\frac{(\widetilde{\mu}_f-\mu_f)^2}{8G_V}\right] \,,
\eeq
where $N_c=3$ is the number of colors and the quark contributions for each flavor are
\begin{align}
\Omega_f&= \Omega_f^{vac}+\Omega_f^{med} \,,
\\
\Omega_f^{vac}&= \frac{1}{4\sqrt{\pi}}\frac{\vert e_f H\vert}{(2\pi)^2}\int^{\infty}_{-\infty} dp_3\int_{1/\Lambda^2}^{\infty}\frac{ds}{s^{3/2}} \left(\sum_{\epsilon}e^{-sE_{f,0}^2}+\sum_{n>0,\zeta,\epsilon}e^{-sE_{f,n}^2}\right) \,, \label{eq:omega2fvac}
 \\
\Omega_f^{med}&= -\frac{\vert e_f H\vert}{2\pi^2}\widetilde{\mu}_f b_f-\frac{\vert e_f H\vert}{8\pi^2}\int^{\infty}_{-\infty} dp_3\sum_{\epsilon}(|E_{f,0}-\widetilde{\mu}_f|-|E_{f,0}|)|_{reg}
\nonumber  \label{eq:omega2fmedium}
\\
&-\frac{\vert e_f H \vert}{4\pi^2}\int^{\infty}_{-\infty} dp_3\sum_{n>0,\zeta}(\widetilde{\mu}_f-E_{f,n})\Theta( \widetilde{\mu}_f-E_{f,n})|_{\epsilon=1}, 
\end{align}
with quark energies given by
\begin{eqnarray}
\label{eq:Equarks}
E_{f,0} = \epsilon\sqrt{\Delta_f^2+p_3^2}+b_f,&\quad  \epsilon=\pm\,,\, n=0 \,, \\
 E_{f,n}= \epsilon\sqrt{\left(\zeta\sqrt{\Delta_f^2+p_3^2}+b_f\right)^2+2\vert e_f H\vert n}, & \quad  \epsilon=\pm \,,\, \zeta=\pm \,,\, n>0 \,,
\end{eqnarray}
where $b_f=\frac{q_f}{2}$, and $n=0,1,2,..$ denotes the Landau levels. Notice that this spectrum exhibits a drastic distinction between the modes of the lowest Landau level (LLL), $n=0$, and the rest, $n>0$. The spectrum is asymmetric about zero for the LLL, while it is symmetric for any $n>0$. The index $\zeta$ is connected to the spin projection, while $\epsilon$ labels particle/antiparticle energies for all $n>0$. This last interpretation is not valid however for the LLL, due to the spectral asymmetry. Note that for $n>0$, the presence of the modulation, $b_f\neq 0$, breaks the spin degeneracy, creating a Zeeman effect in the absence of an anomalous magnetic moment term. The first two terms in the r.h.s. of (\ref{eq:omega2fmedium}) were found using the regularization procedure discussed in \cite{Klimenko2010}, and the vacuum term (\ref{eq:omega2fvac}) was regularized with the help of Schwinger's proper time scheme.

The electron thermodynamic potential is
\beq\label{reg-omega-e}
\Omega_e=\Omega_e^{vac}+\Omega_e^{med}=\frac{1}{4\sqrt{\pi}}\frac{\vert eH\vert}{(2\pi)^2}\int^{\infty}_{-\infty} dp_3\sum_{n\epsilon}d(n)\int_{1/\Lambda^2}^{\infty}\frac{ds}{s^{3/2}}e^{-sE_e^2}-\frac{\vert eH\vert }{4\pi^2}\sum_{n}d(n)\int^{\infty}_{-\infty} dp_3(\mu_e-E_e)\Theta( \mu_e-E_e)|_{\epsilon=1},
\eeq
where the degeneracy factor $d(n)=2-\delta_{n0}$ takes into account the lack of spin degeneracy of the LLL and the modes are given by the well-known spectrum of a free charged fermion in a magnetic field,
\begin{equation}\label{eq:electron}
E_e= \epsilon\sqrt{m_e^2+p_3^2+ 2\vert eH \vert n}, \quad  \epsilon=\pm  \,.
\end{equation} 

To find the expectation values of  $\Delta_f$, $b_f$, and $\tilde\mu_f$ we must solve the set of equations
\beq \label{minimumeq}
\frac{\partial\Omega^{(2f)}}{\partial \Delta_f}=0, \qquad \frac{\partial\Omega^{(2f)}}{\partial b_f}=0,\qquad \frac{\partial\Omega^{(2f)}}{\partial \tilde{\mu}_f}=0 \,, \quad f \in \lbrace u,d \rbrace \,,
\eeq
together with the electrical neutrality condition
\beq
\frac{\partial\Omega^{(2f)}}{\partial \mu_e}=0  \,.
\eeq
The third equation in (\ref{minimumeq}) is equivalent to the equation for the baryon density $3\rho_B=-\partial\Omega^{(2f)}/{\partial \mu}$. Its solution corresponds to a maximum of $\Omega^{(2f)}$ \cite{Koide}.

\begin{figure}
\begin{center}
\includegraphics[width=0.49\textwidth]{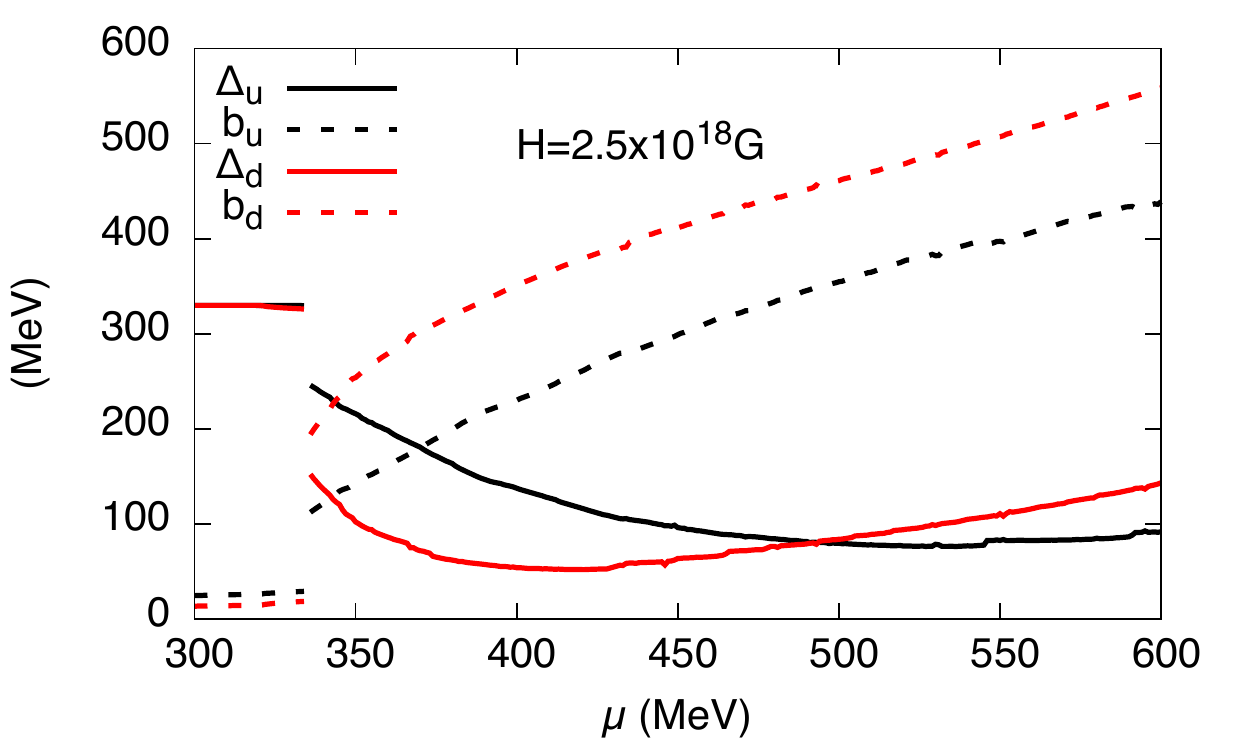}
\caption{ Amplitudes $\Delta_f$ ($f \in (u,d))$ and wave numbers $b_f$ for the CDW modulation (Eq. \ref{condensates2f}) as a function of chemical potential in the presence of a constant external magnetic field. \label{Delta_q_vs_mu}}
\end{center}
\end{figure}

The contribution $-\frac{\vert e_f H\vert}{2\pi^2}\widetilde{\mu}_f b_f$  in (\ref{eq:omega2fmedium}) originates from the asymmetry of the spectrum at the LLL and is directly connected to the baryon charge anomaly \cite{Tatsumi:2014}. Just as in the isospin-symmetric case \cite{Klimenko2010, Tatsumi:2014}, the presence of this term favors nonzero values for $b_f$ for all $\mu > 0$ within the region of validity of the model.  Therefore, strictly speaking, for quark matter in the presence of a magnetic field, the formation of the CDW condensate is energetically favored from arbitrarily small to intermediate values of chemical potentials. This is clearly seen in the plots of Fig. \ref{Delta_q_vs_mu} where the behavior of the magnitude and modulation of the condensates as functions of the baryon quark chemical are depicted for $H=2.5 \times 10^{18}G$. The separation of the parameters $b_f ,\Delta_f$  for each quark is a consequence of the neutrality condition that leads to different quark chemical potentials and hence different condensate solutions for different flavors. However, the behavior of the individual condensates with the chemical potential is similar. For small chemical potentials, the inhomogeneity is present, but the modulation is very small and the magnitudes of the up and down condensates coincide and are equal to the magnitude of the homogenous case. In contrast, in the region of interest for star applications, $330$ MeV$<\mu< 550$ MeV, the two inhomogeneous condensates are quite distinct and robust. At very large chemical potential a competition may occur between the CDW solution and some form of color superconductivity, a topic worth being explored, but out of the scope of this paper. 

\subsection{Three-flavor model}
\noindent
While the exact densities reached in the core of compact stars are unknown, it is likely that strange quarks might play a role in the thermodynamics of these systems. In particular, when building a hybrid EoS, if the transition from nuclear to quark matter occurs beyond the hyperon onset in the nuclear phase, a two-flavor description of quark matter will clearly lead to inconsistencies and the inclusion of a third flavor in our quark model might therefore be necessary.
As in the two-flavor case, imposing the condition of $\beta$-equilibrium leads to a split of the different flavor chemical potentials, which are given by 
\beq
\label{eq:splitmu3f}
\mu_u = \mu - \frac{2}{3} \mu_e \,, \qquad \mu_d = \mu + \frac{1}{3} \mu_e \,, \qquad \mu_s = \mu + \frac{1}{3} \mu_e \,.
\eeq
We then consider a three-flavor version \cite{Klevansky} of the neutral NJL model discussed above,
 with Lagrangian density
\begin{equation} \label{L-ENJL}
\mathcal{L}^{(3f)} =\mathcal{L}_0 +\mathcal{L}_1+\mathcal{L}_V \,,
\end{equation}
with
\begin{equation}\label{L-0}
\mathcal{L}_0 =\bar\psi \left(i \gamma^\mu D_\mu + \hat{\mu} \gamma^{0} - \hat{m}\right) \psi +\bar{\psi}_e \left(i \gamma^\mu D^{(e)}_\mu+\mu_e \gamma^{0}- m_e \right)\psi_e \,,
\end{equation}

\begin{equation}\label{l-13f}
\mathcal{L}_1 =G_S \sum^{8}_{a=0}\left[(\bar\psi\lambda^a\psi)^2 + (\bar\psi i\gamma^5 \lambda^a \psi)^2 \right] \,,
\end{equation}

\begin{equation}\label{l-V3f}
\mathcal{L}_V =-G_V \sum^{8}_{a=0}\left[(\bar\psi\gamma_\mu\lambda^a\psi)^2 + (\bar\psi\gamma^5 \gamma_\mu\lambda^a \psi)^2 \right] \,,
\end{equation}

Here $\hat{m} = {\rm diag}(m_u,m_d,m_s) $ is the current mass matrix (for our calculations we again neglect the light quark masses, while choosing for the strange current mass a value of  $m_s = 150$ MeV), $ \hat{\mu}= {\rm diag}(\mu_u,\mu_d,\mu_s) $ is the flavor chemical potential matrix, $\lambda^a$ are the Gell-Mann matrices in flavor space for $a$ = 1, \dots 8, and $\lambda_0=\sqrt{\frac{2}{3}}\textbf{1}$. The covariant derivatives are defined as before, but with the replacement of the electric charge matrix by  $Q = {\rm diag}\lbrace \frac{2}{3}e , -\frac{1}{3}e,-\frac{1}{3}e \rbrace $ for the quarks.
When choosing our specific ansatz for the mean fields we follow \cite{Moreira:2013ura} and allow only the light quark condensates to become spatially inhomogeneous with the same plane-wave form of \Eq{condensates2f}, while implementing strange quarks as a homogeneous background of quasiparticles with constituent mass $\Delta_s$. Here again, the introduction of a nonzero $\mu_e$ breaks the isospin symmetry in the system, allowing for different values of the quark condensates of different flavors. 
As in the two-flavor case, we introduce quark number densities $\rho_f$ for each flavor, and replace each quark chemical potential by the effective one $\tilde{\mu}_f = \mu_f - 4 G_V \rho_f$.

We have chosen not to include the six-fermion interaction generated by the instanton contribution in the three-flavor case \cite{Klevansky} because for isospin-asymmetric matter, with three condensates and two spatial modulations, it would give rise to coordinate-dependent terms in the mean-field Lagrangian due to the mixing of different condensates like for instance, $\langle\bar\psi_u\psi_u\rangle\langle\bar\psi_d\psi_d\rangle\bar\psi_s\psi_s$, that would make our numerical calculations quite involved. 
 Nevertheless, even though our simplified model is just a first attempt to tackle a very complicated problem, we do not expect that including the instanton term will change the physical picture significantly, as the mass-radius curves should be less sensitive to the degree of mixing than to the main features of the model, namely the asymmetry of the LLL Hamiltonian of the light quarks, and the existence of condensates with different magnitudes and modulations induced by the neutrality condition, all properties that are present whether there is mixing or not.

Working in the mean-field approximation, we readily find that the spectrum of the $u$ and $d$ quarks is still given by Eq. (\ref{eq:Equarks}), thus identical to the two-flavor case. The spectrum of the electrons is the same as before and the energies of the $s$ quarks are given by
\begin{equation}\label{eq:Es}
E_s= \epsilon\sqrt{\Delta_s^2+p_3^2+ 2\vert e_sH \vert n}, \quad  \epsilon=\pm \,.
\end{equation} 

The thermodynamic potential for the three-flavor case is then given by
\beq
\label{eq:omegaiso}
\Omega^{(3f)} = \Omega^{(2f)}+N_c \Omega_s + \frac{(\Delta_s - m_s)^2}{8 G_S}-\frac{(\widetilde{\mu}_s-\mu_s)^2}{8G_V}  \,,
\eeq
where $\Omega^{(2f)}$ is given by \Eq{eq:omega2f} and
\beq\label{reg-omega-s}
\Omega_s=\frac{1}{4\sqrt{\pi}}\frac{\vert e_sH\vert}{(2\pi)^2}\int^{\infty}_{-\infty} dp_3\sum_{n\epsilon}d(n)\int_{1/\Lambda^2}^{\infty}\frac{ds}{s^{3/2}}e^{-sE_s^2}-\frac{\vert e_sH\vert }{4\pi^2}\sum_{n}d(n)\int^{\infty}_{-\infty}dp_3(\widetilde{\mu}_s-E_s)\Theta( \widetilde{\mu}_s-E_s)|_{\epsilon=1}.
\eeq
The dynamical parameters of this model have to be determined from the equations
\begin{eqnarray} \label{eq-3f}
\frac{\partial\Omega^{(3f)}}{\partial \Delta_f}=0, \qquad \frac{\partial\Omega^{(3f)}}{\partial \tilde{\mu}_f}&=&0 \,,\qquad f=u,d,s  \,,
\\
 \frac{\partial\Omega^{(3f)}}{\partial b_f}&=&0 \,, \qquad  f=u,d \,,
\\
\frac{\partial\Omega^{(3f)}}{\partial \mu_e}&=&0 \,.
\end{eqnarray}

Since the s-quark condensate is homogeneous, its role is basically to decrease the number of electrons required for neutrality, but it does not produce any significant qualitative change in the solutions of the inhomogeneous condensates of the light quarks, which display the same behavior with the chemical potential as in the two-flavor case depicted in Fig. \ref{Delta_q_vs_mu}.

\section{Nuclear Matter in a Magnetic Field}

\noindent
For describing nuclear matter we employ the nonlinear Walecka model \cite{Walecka1986}. In the presence of an external magnetic field, it is characterized by the following Lagrangian \cite{Glendenning}:

\begin{equation}
\label{lag_nuc}
\mathcal{L}=\sum_l\mathcal{L}_l+\sum_B \mathcal{L}_B+\mathcal{L}_M
\end{equation}
\\
where
\begin{align}
\mathcal{L}_l&=\bar{\psi_l}(i\gamma_\mu\partial^\mu- e\gamma_\mu A^\mu-m_l)\psi_l \,, \\
\mathcal{L}_B&=\bar\psi_B(i\gamma_\mu\partial^\mu-e_B\gamma_\mu A^\mu-m_B+g_{\sigma B}\sigma -g_{\omega B}\gamma_\mu\omega^\mu-g_{\rho B}\vec{ \tau}\cdot\vec{\rho_\mu}\gamma^\mu)\psi_B \,, \\
\mathcal{L}_M&=\frac12(\partial_\mu\sigma\partial^\mu\sigma-m_\sigma^2\sigma^2)-U(\sigma) + \frac12m_\omega^2\omega_\mu\omega^\mu-\frac14\omega_{\mu\nu}\omega^{\mu\nu}+\frac12 m_\rho^2\vec{\rho_\mu}\cdot\vec{\rho_\mu}-\frac14 \mathbf{\rho}_{\mu\nu}\mathbf{\rho}^{\mu\nu} \,.
\end{align}
The sum is taken over baryons ($B$), considering the nuclear octect (protons, neutrons, and hyperons $\Lambda$, $\Sigma^-$, $\Sigma^+$, $\Sigma^0$, $\Xi^-$, and $\Xi^0$), and leptons ($l$), considering electrons and muons. The mesons ($M$) considered comprise the scalar $\sigma$, isoscalar-vector $\omega_\mu$, and isovector-vector  $\vec{\rho_\mu}$. They mediate the interactions between the baryon Dirac fields $\psi_B$. The lepton Dirac field is represented by $\psi_l$. Here, $e_B$ is the electric charge of each baryon, $\vec{ \tau}=(\tau_1,\tau_2,\tau_3)$ denotes the isospin matrices, and $m_B$ is the baryon mass. 
The field tensors for the mesonic fields are given by  $\omega_{\mu\nu}=\partial_\mu\omega_\nu-\partial_\nu\omega_\mu$ and $\mathbf{\rho}_{\mu\nu}=\partial_\mu\vec{\rho_\nu}-\partial_\nu\vec{\rho_\mu}-g_{\rho B}(\vec{\rho_\mu}\times\vec{\rho_\nu})$, while $U(\sigma)=1/3 b m_n(g_{\sigma N} \sigma)^3 + 1/4 c (g_{\sigma N} \sigma)^4$ is the scalar self-interactions, $m_n$ being the nucleon mass. 

In the mean-field approximation the mesonic fields $\sigma$, $\omega_0$, and $\rho^3_0$ are assumed to acquire nonzero expectation values: $\langle \sigma \rangle=\bar{\sigma}$, $\langle \omega_0 \rangle=\bar{\omega}_0$, and $\langle \rho^3_0 \rangle=\bar{\rho}^3_0$. The mesonic masses are $m_{\sigma}=$ 400 MeV, $m_{\omega}=$ 783 MeV, and $m_{\rho}=$ 770 MeV. The hyperon couplings to the mesonic fields are described as a fraction of that of the nucleons and are taken as $x_{\sigma H}=0.7$ and $x_{\rho H}=x_{\omega H}=0.783$  ($x_{iB}=g_{iB}/g_i$, $i=\sigma, \rho,\omega$). The parameters are chosen to reproduce a binding energy of $-16.3$ MeV and a symmetry energy coefficient of $32.5$ MeV for saturated nuclear matter with compression modulus $K=300$ MeV and effective baryon mass $m_B^{*}=m_B-g_{\sigma}\bar{\sigma}=0.7 m_B$. For this we adopt the following values: $(g_{\sigma}/m_{\sigma})^2$= 11.79 fm$^{-2}$, $(g_{\omega}/m_{\omega})^2$= 7.149 fm$^{-2}$, $(g_{\rho}/m_{\rho})^2$= 4.411 fm$^{-2}$, $b=0.002947$, and $c=-0.001070$ (GM1 parametrization). 

The thermodynamic potential at zero temperature is therefore:
\begin{align}
\label{eq:omegaNucl}
\Omega & = -\sum_{i=B,l}\Omega_i - \frac12\left(\frac{g_{\omega}}{m_{\omega}}\right)^{-2}\rho_B'^2+\frac12\left(\frac{g_{\sigma}}{m_{\sigma}}\right)^{-2}(g_{\sigma}\bar\sigma)^{2}+\frac13 bm_n(g_{\sigma}\bar\sigma)^3+\frac14 c(g_{\sigma}\bar\sigma)^4-\frac12\left(\frac{g_{\rho}}{m_{\rho}}\right)^{-2}\rho_{I_3}'^2  \,,\\
\rho_{I_3}' & = \sum_{i=B}x_{\rho i}I_{3i}\rho_i \,,\\
\rho_{B}' & = \sum_{i=B}x_{\omega i}\rho_i \,,
\end{align}
where the terms for charged particles (with dynamics modified by the filling of the Landau levels) and uncharged particles are given by
\begin{align}
\Omega_i^{neutral}&= -\frac13\frac{\gamma_i}{(2\pi)^3}\int d^3k\frac{k^2}{\sqrt{k^2+m_i^{*2}}} \,,\\
\rho_i^{neutral}&= \frac{k_{fi}^3}{3\pi^2}\,,\\
\Omega_i^{charged}&= -\frac{|eH|}{2\pi^2}\sum_{n=0}^{n_{max}}d(n)\int dk\, \frac{k^2}{\sqrt{k^2+\tilde{m}^{*2}_i}} \,, \\
\rho_i^{charged}&= \frac{|eH|}{2\pi^2}\sum_{n=0}^{n_{max}}d(n)\tilde{k}_{fi} \,,
\end{align}
where $\gamma_i$ is the degeneracy factor and $\rho_i$ the number density. The spin degeneracy of the Landau levels $n$ is denoted as before by $d(n)=2-\delta_{n0}$. The sum is taken from the LLL to $n_{max}$, where $n_{max}$  is the nearest natural number equal to or less than $[(\widetilde{\mu}_i^2-m^{*2}_i)/2|eH|]$ with $\widetilde{\mu}_i=\mu_i-g_\omega\bar{\omega}_0-g_\rho \tau_{3i}\bar{\rho}^3_0$ denoting the effective chemical potential for the given fermion. If we write the effective mass of charged components as {$\tilde{m}^{*2}_i=m^{*2}_i+2n|eH|$},  the Fermi momentum of the particles becomes {$\tilde{k}_{fi}^2=\widetilde{\mu}_i^2-\tilde{m}^{*2}_i$}. The baryon sum can be limited to protons and neutrons (GM1n case) or include the full baryon octet (GM1nh case). The EoS for the two cases will start to differ at densities around  0.3-0.4 fm$^{-3}$ or $\mu_B \sim 1230$ MeV, which mark the onset of hyperons \cite{Baldo, Veronica}.
The thermodynamically favored values for the variational parameters
 in the model are obtained by solving the field equations describing
 the coupling of baryons to the mesons while considering chemical equilibrium and charge neutrality:

\begin{eqnarray}
\left(\frac{m_\sigma}{g_\sigma}\right)^{2}(g_\sigma \bar{\sigma})+bm_n(g_\sigma \bar{\sigma})^2+c(g_\sigma \bar{\sigma})^3 &=& \frac{1}{\pi^2}\sum_{i=B}\frac{g_\sigma}{m_\sigma^2}\int\frac{m^*_B}{\sqrt{k^2+ m^{*2}_B}}k^2 dk\, 
\\
\bar{\omega}_0 &=&\sum_B \frac{g_\omega}{m_\omega^2}\frac{k_{fB}^3}{3\pi^2}\, , 
\\
\bar{\rho}_0 &=& \sum_B \frac{g_\rho}{m_\rho^2}\frac{k_{fB}^3}{3\pi^2}I_{3B}\, , \\
\sum_{i=B,l} e_i \rho_i&=&0\,,
\\
\mu_i&=&B_i\mu_B+e_i\mu_e\, ,
\end{eqnarray}
\\
where $e_i$ and $B_i$ are the electric and baryonic charges of each component, associated with the corresponding electrical and baryonic chemical potentials, $\mu_e$ and $\mu_B$, respectively.

\section{Varying inner magnetic field}

\noindent
Most compact stellar objects are known to have very high values of surface magnetic fields, with white dwarfs in the range of $10^{6}-10^9$G, typical neutron stars $10^8-10^{12}$G, and magnetars $10^{14}-10^{15}$G \cite{magnetar1, *magnetar2, *magnetar3, *magnetar4, *magnetar5}.
While the magnetic field is definitely expected to rise in several orders of magnitude when going from the surface to the core, its actual inner profile is unknown.  Estimates based on the virial theorem for stars made of quark matter give upper values of central fields of order $10^{19}-10^{20}$G \cite{Ferrer2010}. Other estimates found by solving the Einstein equations with axisymmetric and poloidal field configurations \cite{Bocquet, Cardall}, or by applying the virial theorem to stars entirely composed of nuclear matter \cite{Dong}, have led to the lower range $0.1 - 4.2 \times 10^{18}$G. 

In all our derivations, we consider a static background magnetic field $H$ pointing in the $z$-direction. This field influences the EoS both by altering the energy spectrum of the charged particles and by producing a splitting between the parallel and transverse pressures with respect to the field direction \cite {Ferrer2010}. 
The pressure splitting is mostly due to the Maxwell term 
$\mathcal{L}_{\rm EM} = -\frac{1}{4}F^{\mu\nu}F_{\mu\nu}$, where $F_{\mu\nu}=\partial_\mu A_\nu - \partial_\nu A_\mu$, contribution to the Lagrangian. We shall work in a region of magnetic field strengths where the pressure splitting is $\leqslant 10\%$, so that the ambiguity associated with using the spherical TOV equations to obtain the mass-radius sequences is also small. Additionally, we neglect the interaction of the magnetic field with the anomalous magnetic moment of the particles because its inclusion has been shown to have a negligible effect on the EoS of the dense medium \cite{Ferrer2015}.

In the stellar medium, the electric conductivity is very large; thus the magnetic flux is conserved. Hence, the magnetic field should be stronger as the density increases towards the core. Assuming a constant field magnitude throughout the star would then be a very crude approximation, which might introduce a significant bias in the resulting EoS. To avoid this, we consider a varying magnetic field in the star. A first implementation of a varying magnetic field inside the star was done in  Ref. \cite{H_profile} and then used by several authors \cite{Mao03,*Menezes09,*Rabhi09,*Casali14}. There, the neutron star was assumed to be composed entirely of nuclear matter and the magnetic field was expressed as a function of the baryon density, changing from a maximum central value to some lower surface value that was estimated from known magnetars. This ansatz, however, is not convenient to study hybrid stars on which the change from nuclear to quark matter occurs through a first-order transition.  The jump that occurs in the baryon density due to the first-order transition would in turn produce an unphysical jump in the magnetic field. In such a case, a better way to mimic the inner varying field is to express it as a function of the baryon chemical potential, as proposed in \cite{Veronica}. Here we adopt the same approach and consider a medium-dependent value of $H$ when calculating the EoSs of the nuclear and quark phases.  With this aim, we employ the following ansatz:
\begin{equation} \label{varB}
H(\mu_B)=H_{S}+H_C\left[1-e^{ -\kappa \left (\frac{\mu_B-\mu_N}{\mu_N} \right )^\gamma} \right ] \,,
\end{equation}
where  $\mu_N = 938$ MeV can be interpreted as the baryon chemical potential for nuclear matter at the crust of the star. The parameters $\kappa$ and $\gamma$ determine how fast the rise of $H$ is with the baryon chemical potential from the surface to the core. $H_{S}$ is the value for the surface magnetic field and $H_C$ an estimate of the value at the core.  The uncertainty in the knowledge of the actual inner profile of the magnetic field is expressed  in (\ref{varB}) by the arbitrariness of the $(\gamma,\kappa)$ parametrization. While building hybrid stars, in Sec.\ref{sectionV}B we  consider different values within acceptable ranges for these parameters, in order to determine the sensitivity of the EoS and the maximum stellar mass to their variation.

\begin{figure}
\begin{center}
\includegraphics[width=0.49\textwidth]{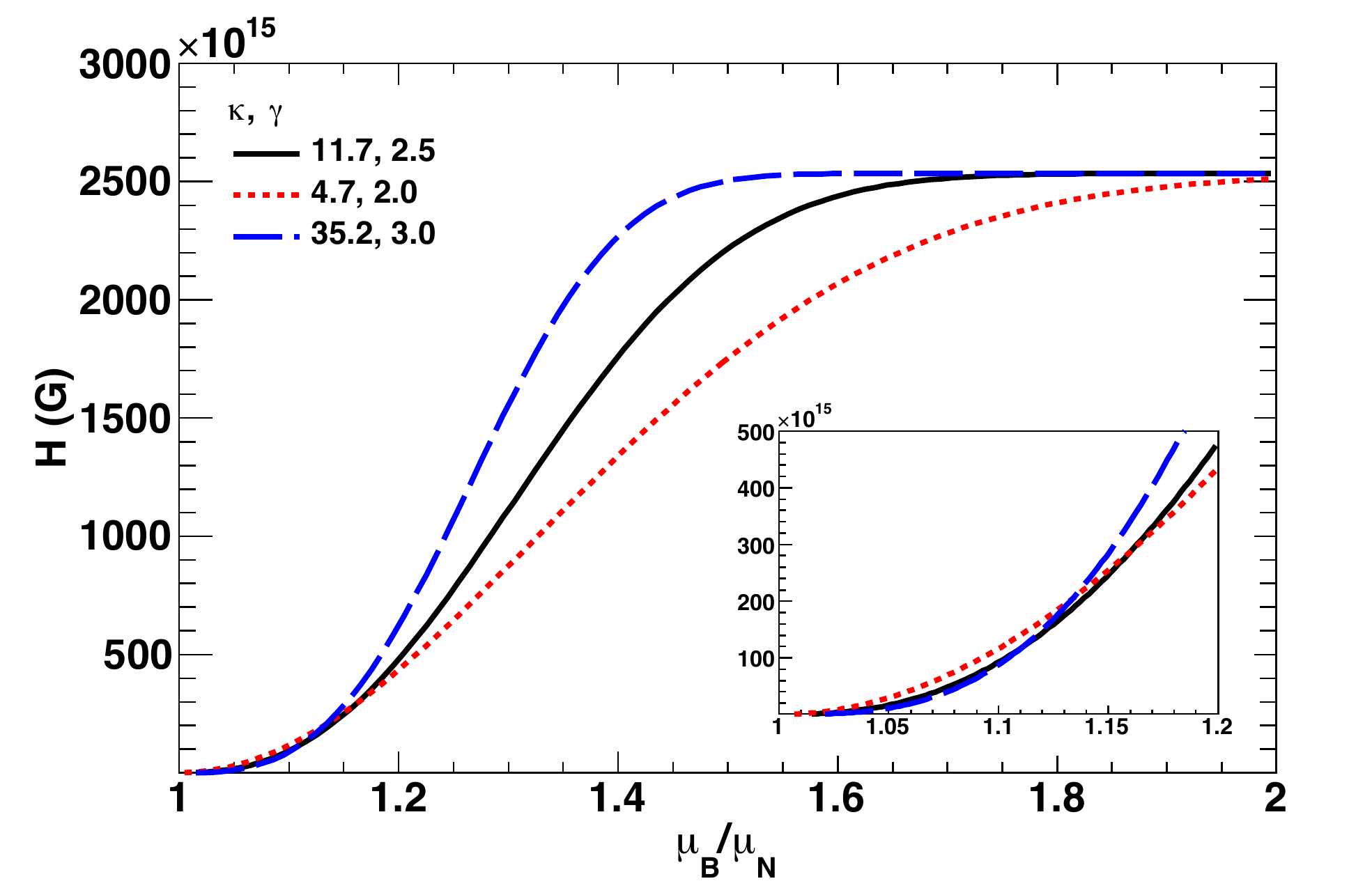}
\caption{ Magnetic field profiles corresponding to Eq. (\ref{varB}) for different parametrizations ($\kappa, \gamma$) in the chemical potential region under consideration.}\label{B-Profile}
\end{center}
\end{figure}

In Fig.\ref{B-Profile} we show the magnetic field profiles for different sets of the $(\kappa, \gamma)$ parameters. Note that for the selected parametrization, the field decays quicker for densities corresponding to nuclear matter, while in the quark core it is almost constant.

\section{Inhomogeneous quark matter in the core of compact stars}\label{sectionV}

\noindent
Since the first indications of the existence of 2$M_{\odot}$ stars \cite{MassStars, MassStars2}, there were claims \cite{Trumper, Ozel} that quark matter might have to be ruled out as a core phase. The problem was motivated by the fact that despite more than a decade of perturbative QCD predicting maximum stellar masses for quark stars larger than 2$M_{\odot}$ \cite{Fraga, Kurkela}, the expected densities of compact stars are not sufficiently high to validate a perturbative approach for the strong interaction. On the other hand, nonperturbative calculations based on simple QCD phenomenological models like NJL with four-fermion channels were found to produce too soft an EOS, incapable to stabilize a 2$M_{\odot}$ star against the gravitational collapse (see for example \cite{Bordbar} and Fig. 8 in \cite{mag3}). However, when other interactions, also present in a dense medium of quarks \cite{Kitazawa2002}, such as the diquark channel \cite{Horvath} and/or the vector channel, were taken into account, the EOS stiffened enough to meet the 2$M_{\odot}$ challenge. Because of these findings, quark matter was back in the competition as a possible core phase of massive stars (see for instance \cite{Orsaria, Menezes:2014, Hell}). One may wonder if this last conclusion could be challenged in turn by the presence of an inhomogeneous quark matter phase in the moderate density region.
 
Exploring the suitability of the CDW phase for the core of neutron stars is particularly important, given that practically all neutron stars have nonzero magnetic fields, and once a magnetic field is present, the CDW phase is known to be energetically favored over the homogeneous chiral phases of quark matter, the chirally broken (at low $\mu$) and the chirally restored (at intermediate $\mu$).  As discussed in Sec. II, the reason why the CDW solution is energetically favored is connected to the anomaly of the baryon charge produced by the spectrum asymmetry of the LLL.  The charge anomaly leads to the term $-\frac{\vert e_f H\vert}{2\pi^2}\widetilde{\mu}_f b_f$ in the thermodynamic potential that increases the pressure, rendering the EoS stiffer. Moreover, from the behavior of the parameters $b_f$ with the quark baryon chemical potential shown in Fig. \ref{Delta_q_vs_mu}, it is evident that this term will become more and more important with increasing $\mu$. What is unclear however is whether this term alone will be enough to compensate for the softening of the EoS 
typically associated to a transition to quark matter, or if other common stiffening factors such as vector interactions will also be needed. The only way to find out is through explicit calculations. Consequently, a main question we aim to explore in this section is the following: Can a magnetized hybrid star with a core of quark matter in the CDW phase sustain a star mass consistent with the $2M_\odot$ observations for an acceptable range of the model parameters? 

\subsection{Pressure splitting and Maxwell construction}

\noindent
To find the transition point from nuclear to quark matter, we employ the Maxwell construction, prescribing that the transition between two phases $1$ and $2$ occurs at the same baryonic chemical potential, $\mu_{B1}=\mu_{B2}$, temperature, $T_1=T_2$, and pressure, $P_1=P_2$. The transition is then of first order and the density exhibits a discontinuity at the phase transition. It was shown in \cite{Debora} that the Gibbs construction, for which the continuity of the electron chemical potential is also required, leads to very similar results for the macroscopic properties of a compact star, so that the choice of the Maxwell construction should be acceptable. 

The inclusion of a background magnetic field can in principle introduce a richer scenario in the construction of a hybrid EoS. Indeed,
as mentioned in the previous section, when a system is subject to an uniform magnetic field along a specific direction, the pressure of the system develops a splitting in the directions parallel and perpendicular to the applied field. This splitting has to be taken into account in the EoS, which is then modified from the usual form into \cite{Ferrer2010}  

\begin{align}
P^{\parallel}& =-\Omega-\frac{H^2}{2} \,,\\
\label{eq:pperp}
P^{\perp} & =-\Omega - H\mathcal{M} + \frac{H^2}{2} \,, \\
\varepsilon & =\Omega+\mu \rho +\frac{H^2}{2} \,, \label{eq:split}
\end{align}
\\
where $\Omega$ is the matter contribution to the thermodynamical potential evaluated at the physical minimum, $\rho=-\partial\Omega/\partial\mu$ is the particle density, $\varepsilon$ the energy density and ${\cal M}=-\partial\Omega/\partial H$ the magnetization.

\begin{figure}
\begin{center}
\includegraphics[width=0.49\textwidth]{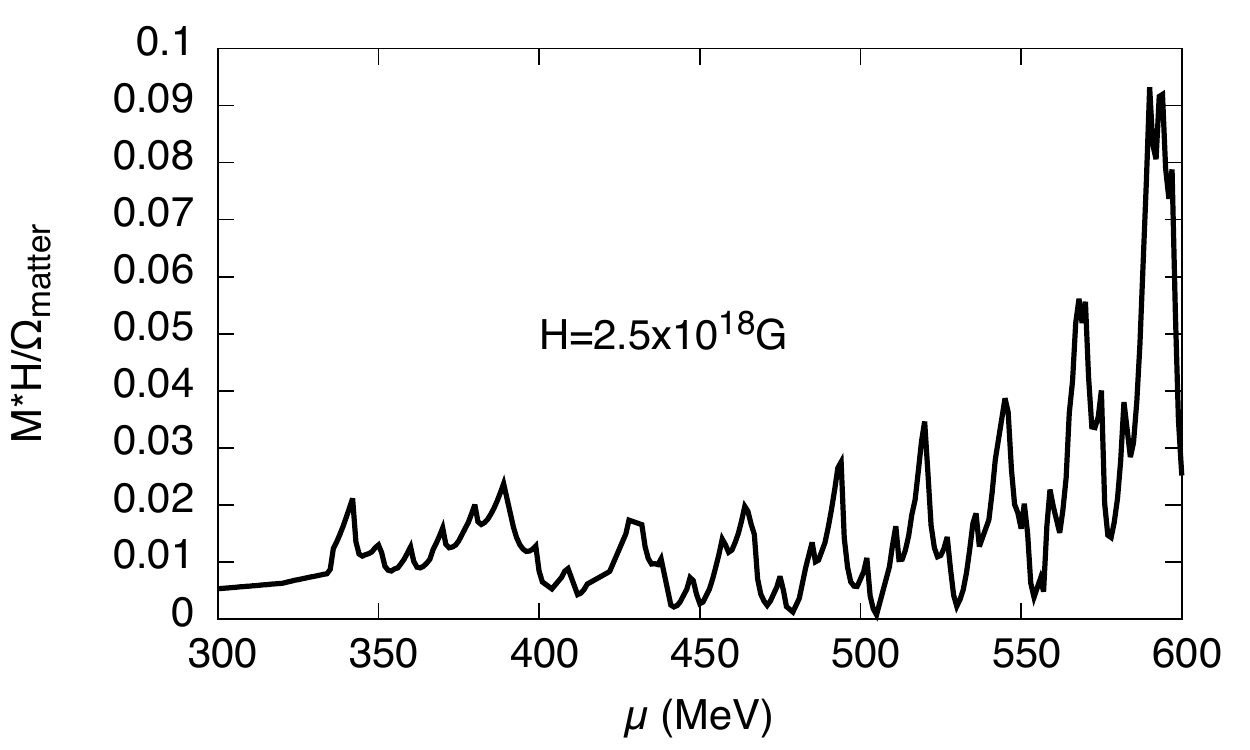}
\caption{Magnetization contribution compared to the matter thermodynamic potential (\Eq{eq:omega2f}) as a function of the quark chemical potential for a two-flavor system in a fixed external magnetic field. The Haas-van Alphen oscillations are richer in the isospin-asymmetric matter here considered due to the difference in the maximum Landau levels reachable for the up and down quark.} \label{Magnetization}
\end{center}
\end{figure}

In light of the pressure anisotropy, the Maxwell construction has to be generalized to require separately the equality of the pressure components of the two phases. Nevertheless, taking into account that the leading term in the thermodynamic potential is $\sim \mu^4$, and that for nonferromagnetic media ${\cal M}<H$, it follows that for the region under consideration, where $H<\mu^2$, neglecting the magnetization energy (${\cal M}H$) in the transverse pressure is  a reasonable assumption. As a corroboration of these arguments, a direct calculation of the magnetization term for quark matter shows that for the range of $\mu$ relevant for star applications $330$ MeV$<\mu< 500$ MeV, $H{\cal M}/\Omega$ never exceeds $4\%$ (see Fig. \ref{Magnetization}), while for nuclear matter $H{\cal M}/H^2$ does not exceed $3\%$ \cite{Broderick2000}. Thus, the magnetization contribution can be neglected in the two phases. This, together with the fact that the magnetic field at the interphase is the same, reduces the Maxwell construction to  

\beq \label{Maxwell-Cond}
\Omega_{\rm nuclear}(\mu_{tr}) = \Omega_{\rm quark}(\mu_{tr}) \,,
\eeq
\\
with $\Omega_{\rm nuclear}$ given by \Eq{eq:omegaNucl} and  $\Omega_{\rm quark}$ given either by \Eq{eq:omega2f} or \Eq{eq:omegaiso}, depending on whether the quark matter is composed of two or three flavors. Using this procedure, the transition chemical potential $\mu_{tr}$ 
can be obtained from the solution of Eq. (\ref{Maxwell-Cond}). 

Upon closer inspection, one can see that this criterion is however not entirely consistent: since the NJL model does not include gluonic degrees of freedom, its pressure will be completely blind to any effect related to confinement, while the nuclear model considered obviously deals exclusively with confined objects. 
In this sense, the transition occurring at $\mu_{tr}$ should be interpreted as a ``deconfinement'' phase transition, which is not built from fundamental gluon dynamics, but only as an effective construction relying on the two phenomenological models involved. While a proper inclusion of confinement properties and the interplay between chiral and deconfinement phase transitions is clearly beyond the scope of this paper, in the spirit of previous works \cite{QM3,Pagliara} we attempt to incorporate these effects in a crude way through the introduction of a constant shift to the NJL vacuum pressure $\delta\Omega_0$, which will be treated as a free parameter. 

\begin{figure}
\begin{center}
\includegraphics[width=0.49\textwidth]{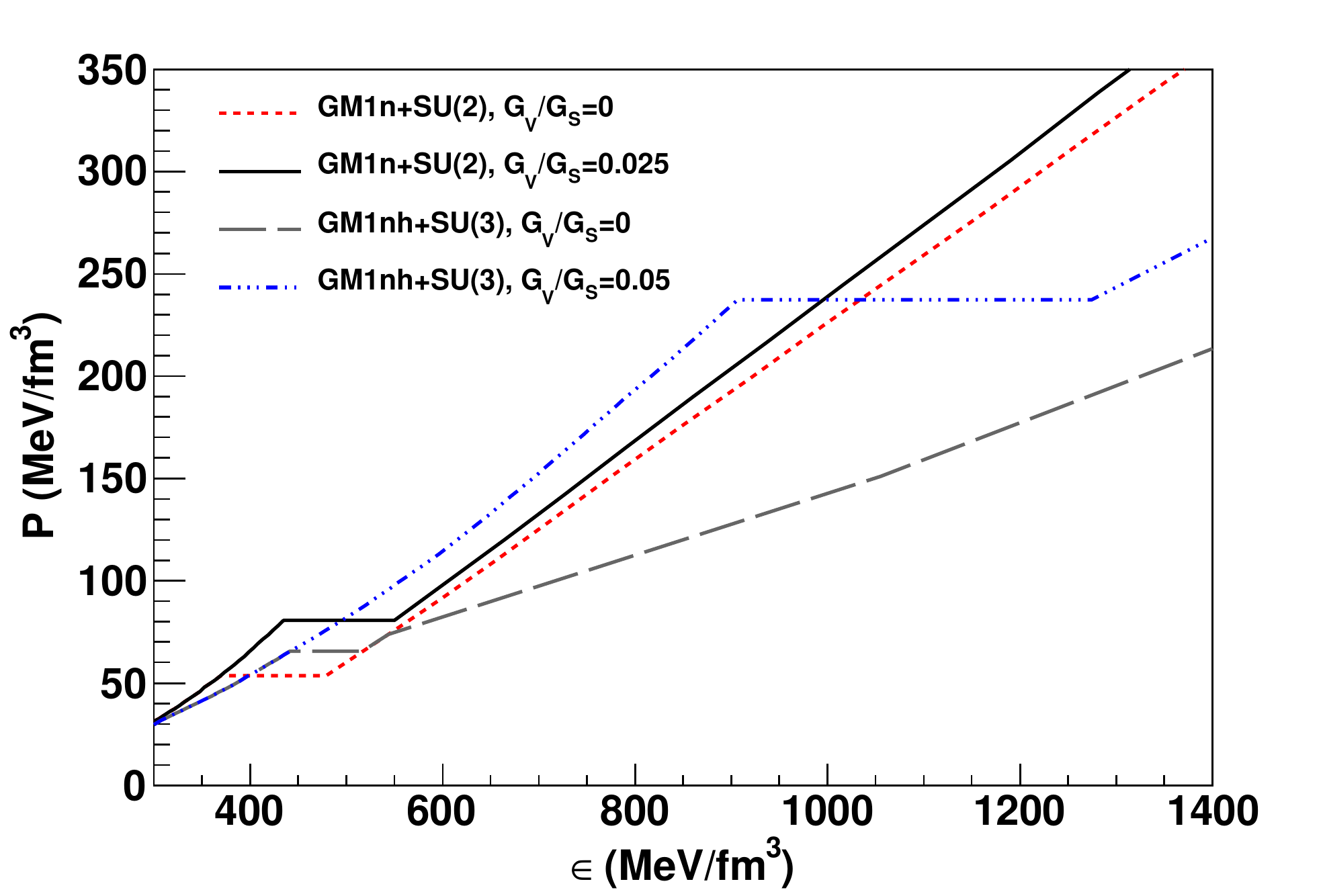}
\includegraphics[width=0.49\textwidth]{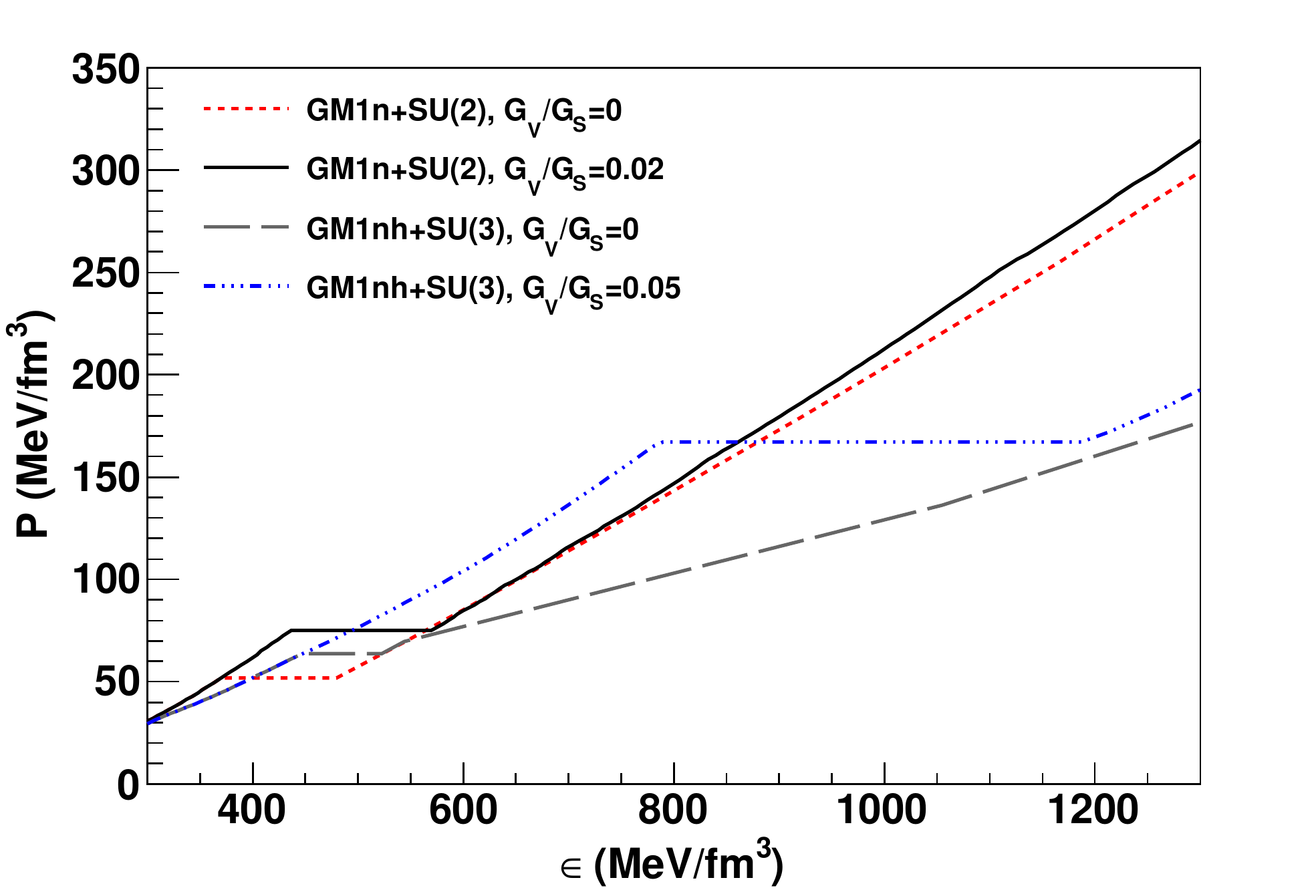}
\caption{Equation of state (considering $P_{\perp}$ on the left panel and $P_{\parallel}$ in the right one) for hybrid stars for different values of the vector repulsion as indicated. The nuclear EoS does not include hyperons when considering quark matter with only quarks up and down (GM1n+SU(2) case) and includes hyperons when the strange quark is included in the quark phase as well (GM1nh+SU(3) case). In the first case, the phase transition must happen at lower values of $\mu_B$. The magnetic field follows \Eq{varB} and is taken with $H_{\rm S}=1\times 10^{15}$ G, $H_{C}=2.5\times 10^{18}$ G, $\gamma=2.5$, and $\kappa=12$. The horizontal lines joining the jumps in the energy density produced by the first-order phase transitions are only graphical artifacts to connect the curves before and after the phase transitions. No shift in the vacuum value of the quark phase was introduced.}\label{Fig_eos_hybrid}
\end{center}
\end{figure}

In the following, as done e.g. in \cite{Debora} for the case of homogeneous condensates, we consider two possible scenarios, one where strange quarks do not contribute to the thermodynamics of the star, and another in which they are included. For the first case  (which we refer to as SU(2) case), we neglect hyperons in the nuclear EoS (GM1n case) and consider only the two light flavors for the quark part. In order to have a consistent SU(2) description, the phase transition to quark matter must occur before the onset of hyperons. This limits the maximum value of the vector coupling in our calculations to $G_V \sim 0.02 G_S$ with $\delta\Omega_0=0$. 
For the SU(3) case, such a limitation is not present, although if $G_V > 0.05 G_S$ the transition chemical potential is so high that a quark matter core is never realized. 
The hybrid EoS for each case, using the parallel and perpendicular pressures, is shown in Fig. \ref{Fig_eos_hybrid}.
We note that the phase transition happening at high chemical potentials would present a much more prominent density jump. It is also noticeable in the curve for GM1nh+SU(3) the softening of the EoS when the strange quark appears, as well as the stiffening of the quark EoS with increasing $G_V$.

\subsection{Masses and radii} 
\label{sec:MR}

\begin{figure}
\begin{center}
\includegraphics[width=0.49\textwidth]{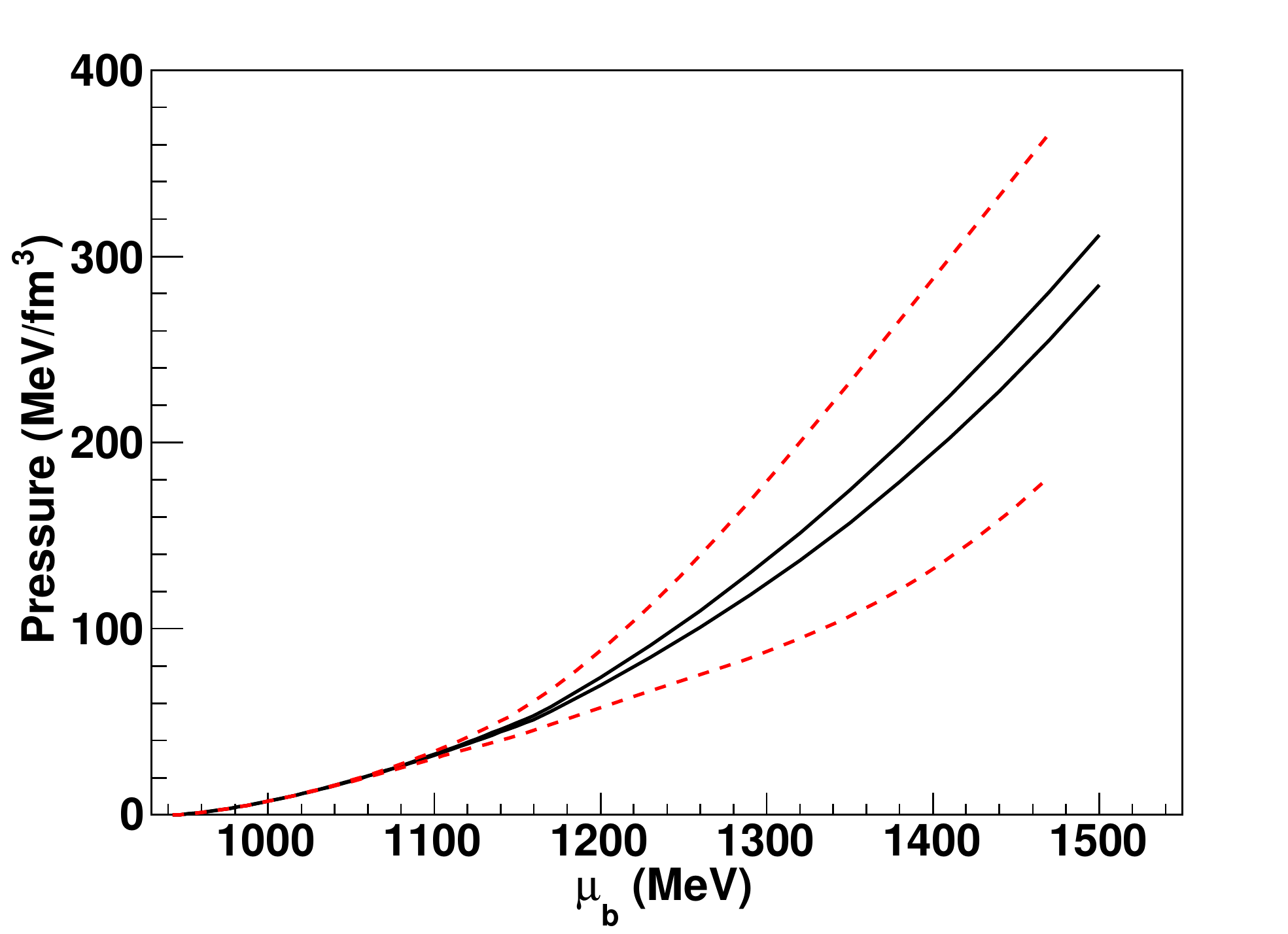}
\caption{Parallel (lower curves) and perpendicular (upper curves) pressures for a two-flavor hybrid star as a function of the baryonic chemical potential.  The surface magnetic field is taken as $1\times 10^{15}$ G and the central field is $2.5\times 10^{18}$ G for the full lines and $6.8\times 10^{18}$ G for the dashed ones with $\gamma=2.5$ and $\kappa=12$. The curves end at the maximum value of the baryon chemical potential achieved in the interior of each configuration.}\label{p_anis}
\end{center}
\end{figure}

\noindent
We now present numerical results for (M-R) sequences using the hybrid EoS obtained in the previous section, together with the Baym-Pethick-Sutherland \cite{BPS} EoS for the star crust.
As discussed above, the presence of strong magnetic fields breaks the star's spherical symmetry and the usual TOV equations for obtaining the star's structure are no longer valid  \cite{mag3}. However, one can consider a range of magnetic fields that is physically meaningful for compact stars and yet does not produce a sizable splitting in the pressure. An example of the pressure splitting profile inside the star is given in Fig. \ref{p_anis}. As expected, such splitting of the pressures is more prominent with an increase in the field, which translates to higher densities given the density dependence of the magnetic field inside the star. One can see from Fig. \ref{p_anis} that if the field remains below $ 3 \times 10^{18} G$ the relative error associated with using one of the two pressures as representative of the whole interior of the star remains relatively small ($\leq 10\%$). A similar scale has been found using the EoS of other models of dense quark matter \cite{mag3}. Therefore, we work within the region of fields satisfying $ H \leq 3 \times 10^{18}$ G, so as not to invalidate the use of the spherical TOV equations. As a cross-check, we perform our calculations using  both pressures, in order to make sure that the choice of one over the other does not lead to dramatic changes in our results.

\begin{figure}
\begin{center}
\includegraphics[width=0.49\textwidth]{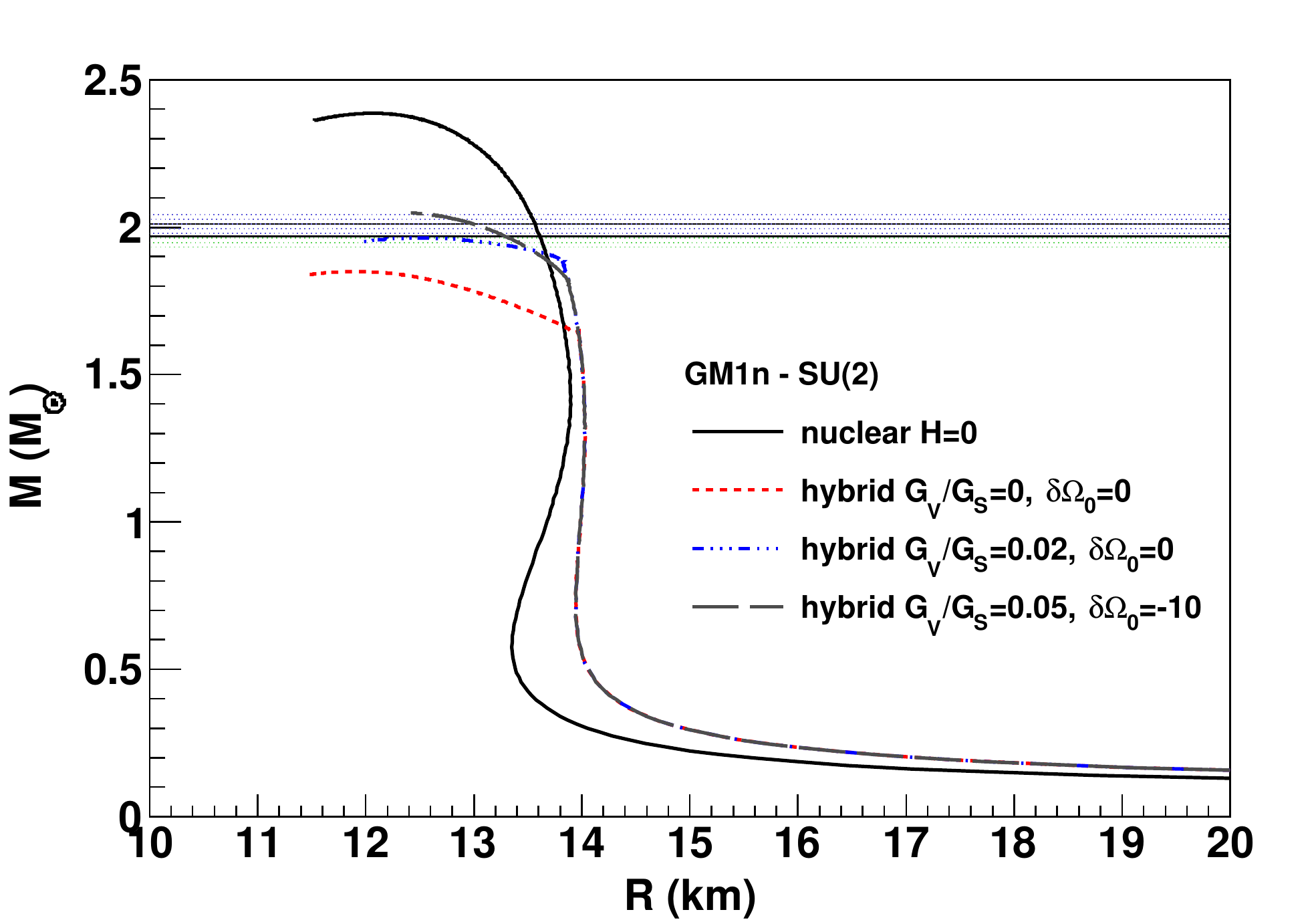}
\includegraphics[width=0.49\textwidth]{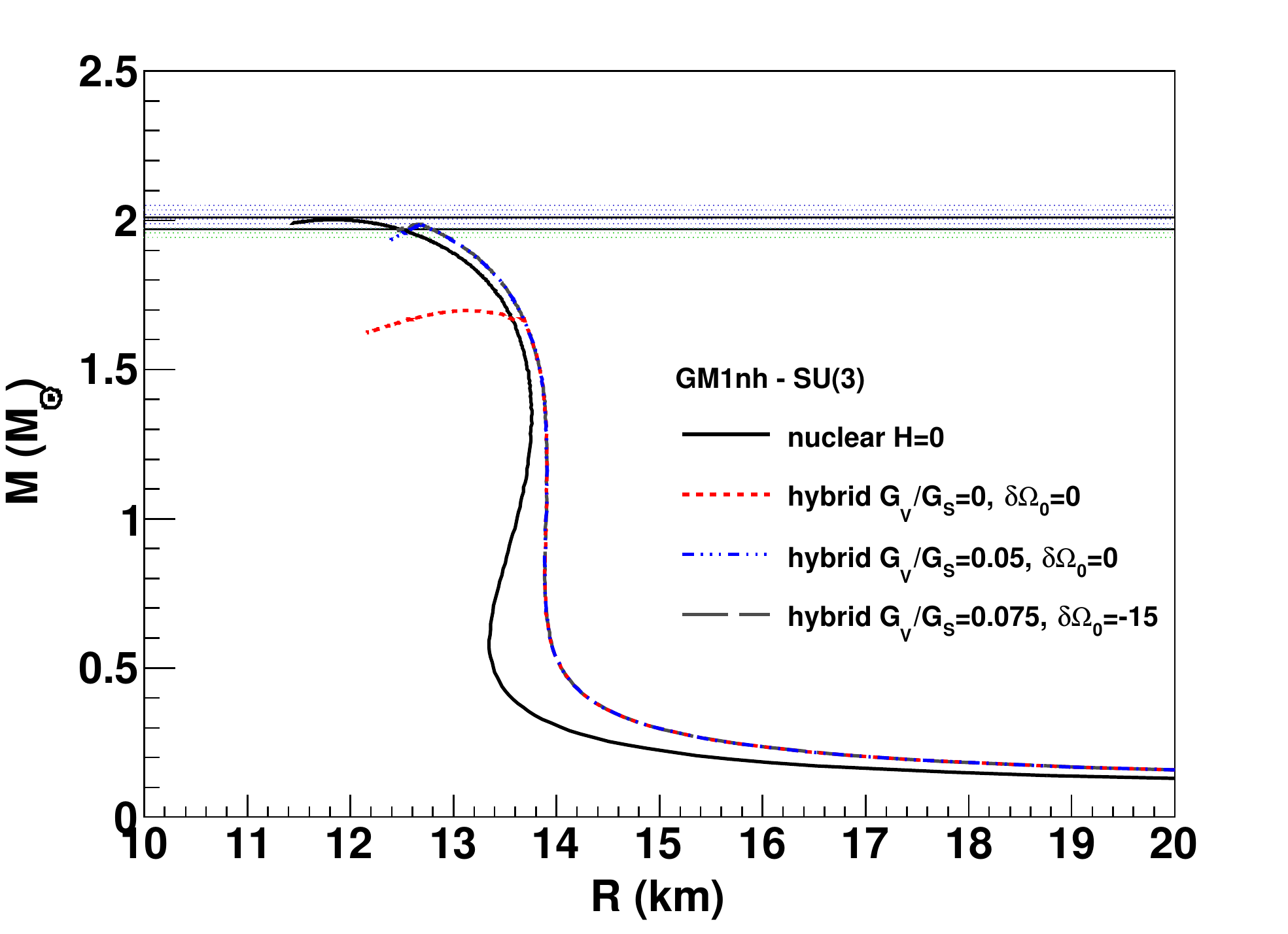}
\caption{The mass-radius relations for a two-flavor (left) and three-flavor (right) inhomogeneous hybrid star considering $H_{S}=1\times 10^{15}$G, $H_C=2.5\times 10^{18}$G, $\gamma=2.5$, and $\kappa=12$. We employed the perpendicular pressure and the shifted vacuum pressure in units of MeV/fm$^3$ (see text for details) is indicated. The curves for stars composed entirely of nuclear matter in the GM1 parametrization with no magnetic field are also shown for comparison. The mass constraints ($\pm \sigma$) of pulsars PSR J1614-2230 and PSR J0348+0432 are shown as shaded regions.}\label{MR_GM1}
\end{center}
\end{figure}

 Hybrid star M-R sequences with an inhomogeneous quark matter core are shown in Fig. \ref{MR_GM1} for values of surface magnetic field compatible with magnetars, $H_{S} \approx 1 \times 10^{15} $G and $H_C \approx 2.5 \times 10^{18}$ G, using the perpendicular pressure.
  From the shape of these curves, the shrinking of the quark matter core with increasing $G_V$ is clearly visible. The effective field in the center of the star is usually smaller than $H_C$ as can be seen in Table \ref{table}. As expected, the inclusion of strangeness for both the nuclear and quark phases softens the EoS, so that the maximum mass is substantially reduced. In the case of GM1nh+SU(3) with $G_V=0.05$, the phase transition happens for a very high value of the central density and the quark core turns out to be extremely small.

As previously mentioned, we also investigated the effects of including a constant shift $ \delta\Omega_0$ in the NJL model vacuum pressure, which we interpret as a contribution from confining effects (i.e. a bag constant). Following \cite{QM3,Pagliara}, this quantity is treated as a free parameter ranging between $-17 \leq \delta\Omega_0 \leq 0 $ MeV/fm$^3$. It is clear from Fig. \ref{MR_GM1} that the effect of the charge anomaly alone is not large enough to push the mass to the desired $2M_\odot$ value. However, choosing a negative value of $\delta\Omega_0$ results in a slightly stiffer EoS for quarks and pushes the phase transition from nuclear to quark matter to lower chemical potentials (albeit always larger than $\mu = 350$ MeV). This allows the use of a larger value of the vector coupling, which will also help make the quark EoS stiffer. The left panel of Fig. \ref{MR_GM1} shows that with the combination of all these effects, for suitable values of $G_V$ and $\delta\Omega_0$, it is possible to increase the maximum mass achieved in the two-flavor case. Although the mass increases only by a relatively small amount, it is enough to make it compatible with the observations of PSR J1614-2230 and PSR J0348+0432.

 In the GM1nh+SU(3) case, a lower value of the transition chemical potential and a stiffer quark phase could also be achieved by the use of a negative shift in the vacuum value (see the right panel of Fig. \ref{MR_GM1}). However, in this case, due to the small size of the quark core, the influence of these effects is greatly reduced compared to the two-flavor scenario. Of course, this is limited by the nuclear EoS used here: different models might allow for larger quark cores and make these effects more noticeable.

\begin{table}
\setlength{\tabcolsep}{6pt}
\begin{tabular}{l| r |r r r r| r| r r}
  \hline
  \hline
 & GM1n ($H$=0) & \multicolumn{4}{c|}{GM1n+SU(2)} & GM1nh ($H$=0) & \multicolumn{2}{c}{GM1nh+SU(3)}\\
\hline
$G_V/G_S$ & - & 0 & 0 & 0 & 0.02 & - & 0 & 0.05 \\
$\gamma$ & - & 2.00 & 2.50 & 3.00 & 2.50 & - & 2.50 & 2.50\\
$\kappa$  & - & 4.70 & 11.70 & 35.20 & 11.70 & - & 11.70 & 11.70 \\
$H_{S}$ ($\times 10^{15}$ G) & 0 & 1.00 & 1-00  & 1.00 & 1.00 & 0 & 1.00 & 1.00 \\
$H_C$ ($\times 10^{18}$ G) & 0 & 2.50 & 2.50 &2.50 & 2.50 & 0 & 2.50 & 2.50\\
$M_{max} (M_{\odot})$ & 2.39 & 1.84 & 1.85 & 1.87 & 1.96 & 2.03 & 1.70 & 1.98\\
$H_{max}/H_C$ & - & 0.78 & 0.94 &  1.00 & 0.93 & - & 0.56 & 0.92\\
  \hline
  \hline
\end{tabular}
\caption{Maximum mass values in units of solar masses obtained using the perpendicular pressure for hybrid stars composed of nuclear matter (GM1) and two-flavor ($GM1n+SU(2)$) and three-flavor ($GM1nh+SU(3)$) quark matter with inhomogeneous condensates for different values of the parameters determining the magnetic field profile inside the star and the quark repulsion strength ($G_V$).}\label{table}
\end{table}

Table \ref{table} shows the maximum masses obtained using different parametrizations for the magnetic field in \Eq{varB}, as well as different values for the vector coupling. The results are obtained considering the perpendicular pressure.
In this case, the maximum masses of the considered magnetized hybrid stars are smaller than those of pure nuclear matter at zero field (GM1n ($H$=0)).

In order to test the dependence of the maximum mass with the magnetic field profile inside the star, we have changed the values of $\gamma$ and $\kappa$ so as to have a faster ($\gamma=3.0$ and $\kappa=35.2$) and a slower ($\gamma=2.0$ and $\kappa=4.7$) increase of the field as one moves toward the center of the star (see Fig. \ref{B-Profile}), while keeping the same values for $H_C$ and $H_S$ for the case $G_V=0$ and two-flavor matter. The difference in the  corresponding maximum masses can be seen in Table \ref{table}. We note that increasing the rate of decay of the field toward the surface, the maximum mass increases very slightly. For the parametrization used, this only amounts to up to $\sim$ 1\% in comparison to the profile with $\gamma=2.5$ and $\kappa=11.7$ used for all other calculations.  Nevertheless, we point out that a more significant difference could be reached by allowing for a very steep increase/decrease of the magnetic field with $\mu$. Whether such a profile would be compatible with the internal structure of compact stars is not known, so in the present work we restricted ourselves to the more conservative parametrizations shown in the table. We also point out that enforcing a larger value of $H_C$ (although excessively strong central fields would lead to large star deformations and invalidate the use of the spherically symmetric TOV equations) is also expected to influence the maximum mass value in a more prominent way.

\begin{figure}
\begin{center}
\includegraphics[width=0.49\textwidth]{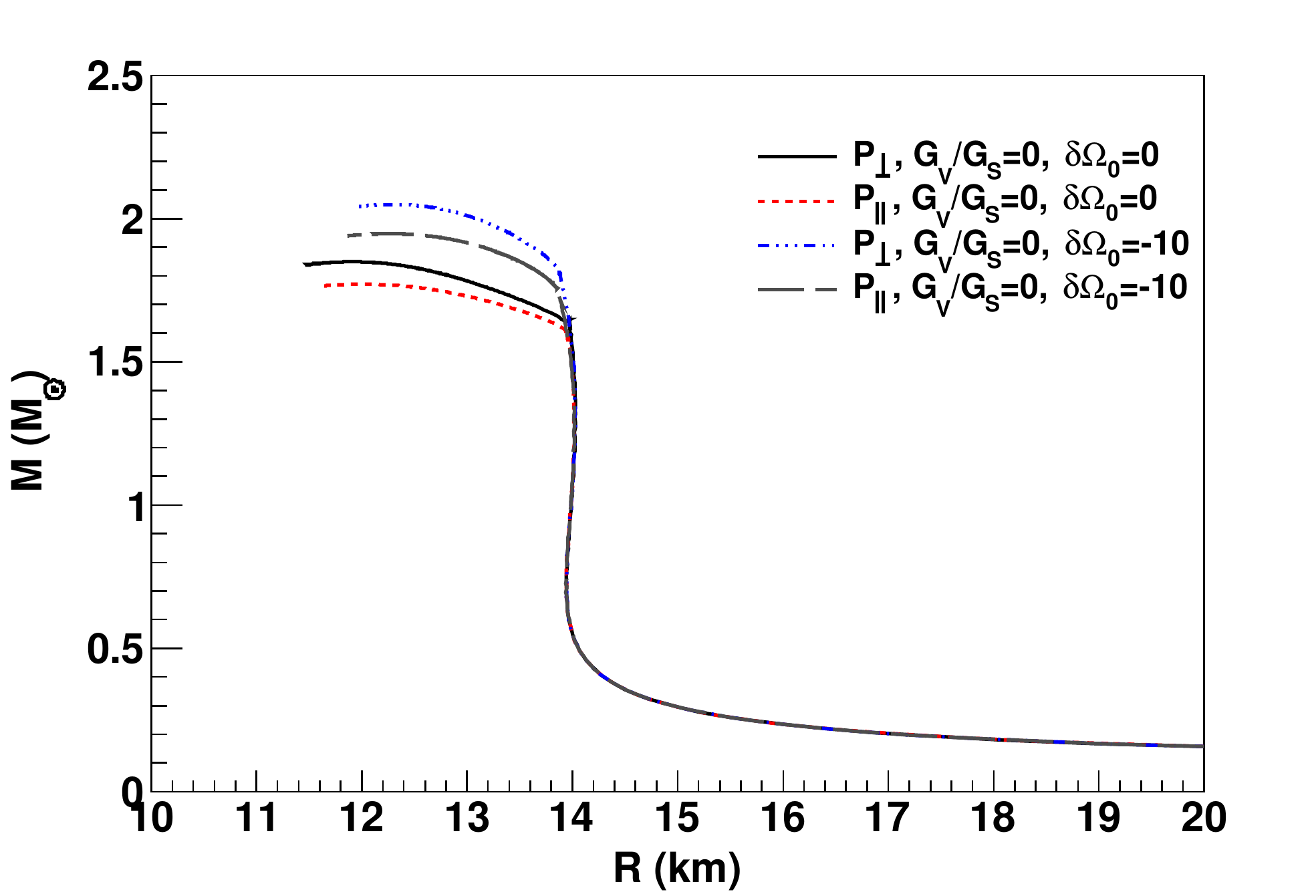}
\caption{ Comparison of the mass-radius relations for a two-flavor hybrid star obtained when using the parallel and perpendicular pressures. The magnetic field is modeled with $H_{S}=1\times 10^{15}$G, $H_C=2.5\times 10^{18}$G, $\gamma=2.5$, and $\kappa=12$. The values of $G_V$ and shifted vacuum pressure in units of MeV/fm$^3$ are indicated. We note that the decrease in the maximum mass attainable is less than 5\% when using the parallel pressure over the perpendicular one.}\label{parXperp}
\end{center}
\end{figure}

The differences arising from the choice of the parallel or perpendicular pressures in the TOV equations are presented in Table \ref{tabledif} and Fig.\ref{parXperp}.
The difference in the maximum masses never exceeds  $0.1$ solar masses and the two curves differ from each other only at high densities, where the magnetic field strength is higher. Unfortunately, given the limitations of the method used here to obtain the star structure in light of the anisotropy introduced by the magnetic field, we cannot conclude which of these results could better represent the maximum mass for a highly magnetized compact object. We therefore simply treat the two results as upper and lower limits of an uncertainty band. In this context, it is interesting to point out that the inclusion of a magnetic field using an axisymmetrical geometry has been found to increase the star mass \cite{Cardall}, a result that might indicate that the actual physical situation in our study could be closer to the upper than the lower limit. Nevertheless, it should also be kept in mind that the conclusions of \cite{Cardall} are not exempt of limitations either, as they were found considering a poloidal field configuration and disregarding the modification of the matter part of the EoS by the magnetic field.

\begin{table}
\setlength{\tabcolsep}{6pt}
\begin{tabular}{l|r r r | r r}
  \hline
  \hline
 & \multicolumn{3}{c|}{GM1n+SU(2)} & \multicolumn{2}{c}{GM1nh+SU(3)}\\
\hline
$G_V/G_S$ & 0 & 0.02 & 0.05 & 0 & 0.05 \\
$\delta\Omega_0$ (MeV/fm$^3$) & 0 & 0 & -10 & 0 & 0 \\
$M_{max}: P_{\perp}$ & 1.85 & 1.96 & 2.05 & 1.70 & 1.98\\
$M_{max}: P_{\parallel}$ & 1.77 & 1.88 & 1.95 & 1.67 & 1.89\\
  \hline
  \hline
\end{tabular}
\caption{Maximum mass (in units of solar mass) allowed for hybrid stars composed of nuclear matter (GM1) and inhomogeneous condensate obtained using the perpendicular and parallel pressures. The varying magnetic field profile is given by $H_S=1\times 10^{15}$ G, $H_C=2.5\times 10^{18}$ G, $\gamma=2.5$, and $\kappa=12$.}\label{tabledif}
\end{table}

Taking all these elements into account, we find that with a relatively low value of $G_V$ and a realistic value for $H_C$, well within the range where the pressure anisotropy is small enough to justify the application of the usual spherically symmetric TOV equations, it is possible to achieve maximum masses around 2 $M_{\odot}$, a result compatible with the precise mass measurements of PSR J1614-2230 ($M=1.97\pm 0.04M_{\odot}$ \cite{Demorest}) and PSR J0348+0432 ($M=2.01\pm 0.04M_{\odot}$ \cite{Antoniadis}). A quark matter core characterized by an inhomogeneous chiral condensate can thus be seen as a viable internal composition of these compact objects.

\section{Concluding Remarks}

\noindent
We studied the effects of the formation of inhomogeneous chiral symmetry breaking phases on the EoS of quark matter in a magnetic field, and the consequences for the  masses of hybrid stars.  To describe quark matter in the core, we considered a conventional NJL model with scalar and pseudoscalar four-fermion channels and included several elements relevant for astrophysical scenarios, such as electrical neutrality, $\beta-$-equilibrium and vector repulsion.  Considering a CDW ansatz for each light flavor, we saw that the spectral asymmetry of the LLL gives rise to a term in the thermodynamic potential that favors the formation of this inhomogeneous ground state and stiffens the pressure. Using this quark model together with the well-established nonlinear Walecka model, we built hybrid EoSs for stars with quark matter core in the CDW phase, and showed that such a configuration can support masses of $\sim 2 M_\odot$. We investigated the sensitivity of these results on the parametrizations involved, and found that masses compatible with the recent PSR J1614-2230 and PSR J0348+0432 measurements can be obtained with reasonable values of the parameters chosen, although in the case where strangeness is allowed, the quark core would end up being significantly reduced in order to achieve higher masses. 

We find the fact that the realization of inhomogeneous quark matter in the core of compact stellar objects gives compatible results with the current mass observation a very encouraging outcome of our investigation, even when considering the still unsolved issue with the splitting of the pressures in the presence of a magnetic field. In light of this, it would definitely be interesting to investigate other effects of the formation of inhomogeneous condensates on the physics of compact stellar objects, such as transport and cooling properties, with the hope of finding stronger experimental signatures.

One may wonder if the  CDW condensate of the magnetized quark system could be washed out by fluctuations. Let us recall that in one-dimensional systems, the Mermin-Wagner theorem \cite{MW,*Coleman} forbids the existence of (1+1)-dimensional Nambu-Goldstone bosons, so that in such systems, condensate solutions that break chiral symmetry cannot be stable against quantum fluctuations about the condensate. In addition, single-modulated condensates in (3+1)-dimensional systems can be quite sensitive to temperature fluctuations because their pairing dynamics is often essentially one-dimensional. This possibility was recently investigated in \cite{Lee:2015bva}, where it was found that the Nambu-Goldstone modes associated with the CDW condensate wash out the long-range order at finite temperatures, although they do support algebraically decaying long-range correlations, so the phase can still exhibit a quasi-one-dimensional order as in liquid crystals. 

While the analysis of our paper has not included the effects of fluctuations, we believe that there are good reasons to expect that in the case with magnetic field, the fluctuations will likely not be as effective in inducing the instability of the CDW condensate. The magnetic field is known to enhance chiral symmetry breaking through the dimensional reduction of the LLL fermions, the well-known mechanism of magnetic catalysis (see \cite{Miransky:2015} for a recent review). The same dimensional reduction has proven to be essential to make the inhomogeneous CDW condensate energetically favored thanks to the asymmetry of the LLL spectrum in the inhomogeneous background. 
Now, the magnetic catalysis mechanism was recently questioned by claims that the dimensional reduction of the LLL fermions would translate into an effective dimensional reduction of the Nambu-Goldstone bosons, which in turn would hinder the stability of the chiral condensate, the so-called inverse magnetic catalysis \cite{Fuku:2013}. But these claims were later challenged by the results of Ref. \cite{MC-KK} that used a functional renormalization group approach to demonstrate that the constituent quark mass increases with the magnetic field at all temperatures and concluded that despite a strong anisotropy in the meson propagation, their fluctuations do not lead to the inverse magnetic catalysis claimed in \cite{Fuku:2013}. In view of this, we expect that a similar behavior to the one found in \cite{MC-KK} should occur in the case of the inhomogeneous chiral condensate. At this point however, we admit that ours are just hand-waving arguments that can be corroborated only by a thorough study of the fluctuations in the CDW phase in a magnetic field, an interesting task to be undertaken in the future. 
 
We remark that in order to obtain a first insight on the effect of inhomogeneous quark matter on the stellar EoS, several simplifying assumptions were made when building our model. One was to ignore the generation of the condensate associated with the anomalous magnetic moment of the chiral pairs. Such a condensate has been found in the presence of a magnetic field for the homogeneous background \cite{Quiroz} and even in the case of color superconductivity in a magnetic field \cite{MCFL-1, LN871, MCFL-2}. There is no reason to expect it is not present also in the inhomogeneous case. However, the magnitude of the magnetic moment condensate is typically small compared to the chiral condensate, except for extremely large magnetic fields,  hence, as a first approximation it can be neglected.  Another simplification was the omission of color-superconducting phases, which are expected to be the true ground state at asymptotically high densities. While we recall that the presence of a magnetic field has been shown to favor the formation of chiral crystalline phases, it is also known that a magnetic field leads to extra condensates in color superconductivity \cite{MCFL-1, MCFL-2}, and makes the MCFL more stable than the regular CFL, so it is expected that a competition between color superconductivity and CDW may occur at intermediate densities in the presence of a magnetic field. Therefore, it is a pending and important task to explore whether the inhomogeneous phase can push or not the onset of color superconductivity in the presence of a magnetic field to densities higher than the ones considered in the present work. An explicit model calculation to address this question would be of course highly desirable. Additionally, an interesting question to consider is whether the incorporation of gluon effects through an extension of the NJL quark model to a gauged NJL quark model could lead to any sizable softening of the EoS of the system, as was recently found to occur in the case of CFL color superconductivity \cite{Laura15}.

Finally, we recall that the inhomogeneous phase considered in this work is just one of the possible exotic phases that could be realized in dense matter. Another plausible candidate for the ground state of strongly interacting matter at intermediate densities, the so-called quarkyonic matter \cite{Mcl}, is also characterized by a spatially varying chiral condensate \cite{Kojo, InceraNJLquark}, although possibly with very different characteristics, particularly in a magnetic field background \cite{Ferrer:2012}, which could result in new unexpected effects on the EoS and on the transport properties of cold and dense quark matter.

\begin{acknowledgments}
S.C. and L.P. are grateful to M. Chiapparini for very helpful discussions. The work of E.J.F. and V.I. has been supported in part by DOE Nuclear Theory grant DE-SC0002179. L.P. acknowledges the financial support received from the Brazilian funding agencies CNPq, Conselho Nacional de Desenvolvimento Cient\'ifico e Tecnol\'ogico, and Fapesp, Funda\c c\~ao de Amparo \`a Pesquisa do Estado de S\~ao Paulo (2013/26258-4), and the hospitality of the UTEP Physics Department where this work was conducted.
\end{acknowledgments}

\bibliography{Inos_final.bib}

\end{document}